\documentclass[journal=jacsat,manuscript=article]{achemso}

\usepackage[version=3]{mhchem} 
\usepackage{braket}
\usepackage{mathtools}
\usepackage{amsthm, amssymb, amsfonts}
\usepackage{cancel}
\usepackage{array}
\usepackage{indentfirst}
\usepackage{threeparttable}
\usepackage[hidelinks]{hyperref}

\newcommand{\ketCC}{\left| \mathrm{CC} \right\rangle }

\author{Xiang Yuan}
\affiliation[Universite de Lille, CNRS, UMR 8523 − PhLAM]
{Physique des Lasers Atomes et Molecules, Universite de Lille, F-59000 Lille, France}
\alsoaffiliation[Vrije Universiteit Amsterdam]
{Department of Chemistry and Pharmaceutical Sciences, Faculty of Science, Vrije Universiteit Amsterdam, 1081 HV Amsterdam, The Netherlands}

\author{Loïc Halbert}
\affiliation[Universite de Lille, CNRS, UMR 8523 − PhLAM]
{Physique des Lasers Atomes et Molecules, Universite de Lille, F-59000 Lille, France}

\author{Lucas Visscher}
\email{l.visscher@ vu.nl}
\affiliation[Vrije Universiteit Amsterdam]
{Department of Chemistry and Pharmaceutical Sciences, Faculty of Science, Vrije Universiteit Amsterdam, 1081 HV Amsterdam, The Netherlands}

\author{André Severo Pereira Gomes}
\email{andre.gomes@univ-lille.fr}
\affiliation[Universite de Lille CNRS, UMR 8523 − PhLAM]
{Physique des Lasers Atomes et Molecules, Universite de Lille, F-59000 Lille, France}

\title[An \textsf{achemso} demo]
  {Frequency-Dependent Quadratic Response Properties and Two-photon Absorption from Relativistic Equation-of-Motion Coupled Cluster Theory}

\abbreviations{IR,NMR,UV}

\begin{document}

\begin{abstract}

We present the implementation of quadratic response theory based upon the relativistic equation-of-motion coupled cluster method. We showcase our implementation, whose generality allows us to consider both time-dependent and time-independent electric and magnetic perturbations, by considering the static and frequency-dependent hyperpolarizability of hydrogen halides (HX, X = F-At), providing a comprehensive insight into their electronic response characteristics. Additionally, we evaluated the Verdet constant for noble gases Xe and Rn, and discussed the relative importance of relativistic and electron correlation effects for these magneto-optical properties. Finally, we calculate the two-photon absorption cross-sections of transition ($ns^{1}S_{0}\to (n+1)s^{1}S_{0}$) of Ga$^{+}$, and In$^{+}$, which are suggested as candidates for new ion clocks. As our implementation allows for the use of non-relativistic Hamiltonians as well, we have compared our EOM-QRCC results to the QR-CC implementation in the DALTON code, and show that the differences between CC and EOMCC response are in general smaller than 5\% for the properties considered. Collectively, the results underscore the versatility of our implementation and its potential as a benchmark tool for other approximated models such as density functional theory for higher-order properties.

\end{abstract}

\section{Introduction}

Nonlinear optical properties (NLO) of matter provide a wealth of information on intra- and inter-molecular interactions and are therefore widely studied in science and engineering\cite{cronstrand_multi-photon_2005,papadopoulos_non-linear_2006,barron_molecular_2009}. NLO properties are also central to materials and device design, with numerous important applications such as optical devices for data transfer and storage. Among the materials being considered, there is a growing interest in NLO properties of molecules containing heavier elements, particularly in Lanthanide\cite{senechal_first_2004,law_nonlinear_2010,andraud_lanthanide_2009,tancrez_lanthanide_2005,senechal-david_synthesis_2006,valore_fluorinated_2010} and Actinide\cite{klepov_trisacrylatouranylates_2014,wang_polarity_2010,serezhkin_syntheses_2017} complexes, as they can offer a superior performance compared to molecules that contain only light elements. 

In order to compute and analyze molecular properties in the linear and non-linear regime, one typically resorts to response theory\cite{christiansen1998response,helgaker_recent_2012,norman_perspective_2011,norman_principles_2018}. Within this theory, the first-order nonlinear response is characterized by the quadratic response function. Quadratic response functions have been implemented for Hartree-Fock (HF) wave-functions\cite{sekino_frequency_1986,rice_frequency_1990} as well as at the electron correlated level employing second-order Møller–Plesset perturbation (MP2)\cite{rice_calculation_1992}, multiconfigurational self-consistent field (MCSCF)\cite{olsen_linear_1985,hettema_quadratic_1992}, coupled cluster (CC)\cite{rozyczko_frequency_1997,hattig_frequency-dependent_1997,gauss_triple_1998}, and density functional theory (DFT)\cite{salek_density-functional_2002,schipper_molecular_2000} reference states. The common starting point of these developments has been the non-relativistic molecular Hamiltonian.

For property calculations, the spin-orbit coupling operator can be added as one of the perturbing properties. That will provide accurate results for lighter elements, at the expense of needing to go one order higher in the responses that are considered. As we move down the periodic table, we reach a point where relativistic effects are too strong to be reliably treated as perturbations. In this domain, it is therefore necessary to refine these methods, ensuring relativistic effects are intrinsically accounted for by employing a variationally stable relativistic Hamiltonian.

In the domain of relativistic quantum chemistry, to date, quadratic response function derivations and implementations are primarily based on mean-field models, such as HF\cite{norman_quadratic_2004} and DFT\cite{henriksson_quadratic_2008}. To improve precision and establish benchmarks for other models, in this manuscript we discuss the development of quadratic response theory based on a relativistic equation-of-motion (EOM)\cite{stanton_equation_1993,liu_relativistic_2017,shee_equation--motion_2018,krylov_equation--motion_2008,bartlett_coupledcluster_2012,sneskov_excited_2012,liu_relativistic_2021} coupled cluster formulation (EOM-QRCC).

We showcase the generality and versatility of our implementation by examining two molecular properties. First, we study the frequency-(in)dependent electric first hyperpolarizability ($\beta$) as it can describe the nonlinear response of a molecule to an applied electric field, which is significant for second-harmonic generation\cite{burland_second-order_1994} associated with the design of optoelectronic devices and can provide valuable insights into the intermolecular interaction\cite{shelton_measurements_1994}. For instance, as discussed by \citet{datta_dipolar_2006}\, $\beta$ is related to the weak intermolecular forces such as dipolar interactions and hydrogen-bonding, thus it is possible to control $\beta$ by modifying the interactions and accurate calculations would be instrumental to provide insight into designing NLO materials like $\pi$-conjugated molecular assemblies.

We consider magnetic circular birefringence, also known as the Faraday effect, as the second property. One example of the interest in studying the Faraday effect can be found in the observation by \citet{savukov_optical_2006} of the inverse Faraday effect in the nuclear magnetic resonance (NMR) sample of liquid water and liquid $^{129}$Xe , which has led to the suggestion that the nuclear spin-induced optical rotation (NSOR) can provide a viable and potentially more informative analog to the NMR chemical shift of traditional NMR detection. There have been only a handful of theoretical investigations of this property, however. For $^{129}$Xe, \citet{ikalainen_fully_2012} performed non-relativistic (NR) time-dependent Hartree-Fock (TDHF), time-dependent Density Functional Theory (TDDFT), coupled cluster response, and relativistic TDHF, TDDFT calculations on the Verdet constant and NSOR. In subsequent work, \citet{cadene_circular_2015} investigated the Verdet constant of $^{129}$Xe in both gas-phase experiments and calculations derived from non-relativistic coupled cluster quadratic response calculations (QR-CC), in which the relativistic effects were approximately accounted for by employing relativistic effective core potentials (ECPs). With our implementation, we shall complement these studies and in particular investigate the relative importance of relativistic (scalar and spin-orbit coupling) effects and electron correlation to these properties.

The characterization of Two-Photon Absorption (TPA) cross-sections, which can be related to quadratic response theory, has also gained considerable attention in different domains and is the third focus of our applications. TPA was first predicted, using perturbation theory, by \citet{goppert-mayer_uber_1931} in 1931, but not observed in experiments until the advent of the lasers that are capable of delivering sufficiently high intensity. The main feature of TPA is that it occurs with a probability depending quadratically on the incident light intensity, which results in the TPA-based techniques offering better spatial resolution than those based on one-photon absorption (OPA). In materials science, materials with large TPA cross-sections enable applications including drug delivery, photodynamic therapy, high-resolution, and optical storage\cite{lee_giant_2021}. Moreover, TPA spectroscopy is also very useful as a research tool. Concerning the different selection rules of TPA compared to OPA, TPA can characterize the excited state in the spectrum in the case of OPA spectrum has been large dispersions, particularly for complex molecules containing f-elements\cite{barker_applications_1987,barker_applications_1992}.

TPA is proportional to the imaginary part of the second-order hyperpolarizability $\gamma$, which requires evaluation of the cubic response function. However, under resonant conditions, it becomes possible to express the TPA cross-sections in terms of the two-photon matrix\cite{hattig_multiphoton_1998}, which can be obtained from the quadratic response of the reference state wave function. With this strategy, the TPA cross-sections have been evaluated in various standard models in quantum chemistry including Hartree-Fock\cite{hettema_quadratic_1992}, MCSCF\cite{hettema_quadratic_1992}, DFT\cite{salek_calculations_2003,frediani_two-photon_2005,nayyar_comparison_2013}, and CC\cite{hattig_coupled_1998,friese_calculation_2012,nanda_two-photon_2015}. Moreover, in the last decades resonant inelastic X-ray Scattering (RIXS)\cite{de_groot_high-resolution_2001,ament_resonant_2011}, a two-photon scattering process involving core electrons, received considerable attention because of the corresponding improvements in sensitivity and energy resolution\cite{fuchs_isotope_2008,hennies_resonant_2010,kunnus_viewing_2016,kjellsson_resonant_2020}, which provides valuable information on the electronic structure of both occupied and virtual states that are not easily accessible by the traditional spectroscopies. Several approaches aimed at the description of RIXS spectra for molecular systems based on non-relativistic or approximate relativistic Hamiltonians have been proposed including algebraic diagrammatic construction (ADC)\cite{rehn_resonant_2017}, MCSCF\cite{josefsson_ab_2012}, DFT\cite{nascimento_resonant_2021}, and EOM-CC\cite{faber_resonant_2019,schnack-petersen_new_2023,skomorowski_feshbachfano_2021,skomorowski_feshbachfano_2021-1,skeidsvoll_simulating_2022,ranga_corevalence_2021}. However, in the relativistic quantum chemistry field, the implementations of TPA cross-sections are still scarce, owing to the additional complexity of handling spin-orbit effects. An implementation in the DIRAC program by \citet{henriksson_two-photon_2005} enabled pioneering calculations of TPA cross-sections from the four-component Hartree-Fock quadratic response theory. In this manuscript, we will focus on TPA for valence processes and will investigate processes involving core electrons such as RIXS in a subsequent publication.

Finally, we pay attention to methods that can lower computational costs. This is of practical importance here since we utilize uncontracted basis sets with adding many diffuse functions, which generate a large virtual orbital space in CC calculations. The simplest and most often used method is the utilization of the MP2 frozen natural orbitals (FNOs)\cite{taube_frozen_2005,taube_frozen_2008,crawford_reducedscaling_2019}. While some authors have pointed out the shortcomings of MP2FNOs for the calculation of linear response properties\cite{kumar_frozen_2017,crawford_reducedscaling_2019}, \citet{surjuse_low-cost_2022} recently suggested using MP2FNOs in EOM-CC calculations can bring about reduce computational cost while retaining sufficient accuracy for ionization energies. On the other hand, to the best of our knowledge, there is no reference yet reporting the performance of MP2FNOs on TPA calculations.

This manuscript is organized as follows: In Sec. 2, the EOM-CC quadratic response theory and the corresponding two-photon absorption matrix formulation are summarized. Section 3 is devoted to the details of the computations we used to test the implementation. The calculations are presented and discussed in Secs. 4. Finally, a brief summary of our findings is given in Sec. 5.

\section{Theory}
    
We base the theory on the time-averaged quasi-energy formalism, which has been summarized
in the landmark paper by~\citet{christiansen1998response}. As the significant part of the formalism to obtain the quadratic response functions is common to that of linear response functions, and we have recently provided an extensive discussion of the implementation details for linear response properties\cite{yuan_formulation_2023}, in the current manuscript, we only focus on the equations related to quadratic response. 

The CC quadratic response function is expressed below:
\begin{equation}
    \begin{split}
        \langle\langle X;Y,Z\rangle\rangle_{\omega_{Y},\omega_{Z}} =& \frac{1}{2}C^{\omega}P^{X,Y,Z}\\
        &\Big[\big[\frac{1}{2}\bold{F}^{X}+\frac{1}{6}\bold{G}\bold{t}^{X}(\omega_{X})\big]\bold{t}^{Y}(\omega_{Y})\\
        &+\bar{\bold{t}}^{X}(\omega_{X})\big[\bold{A}^{Y}+\frac{1}{2}\bold{B}\bold{t}^{Y}(\omega_{Y})\big]\Big]\bold{t}^{Z}(\omega_{Z})
    \end{split}
\end{equation}
in which the wave function is parametrized by the CC amplitudes $\mathbf{t}$ and $P^{X,Y,Z}$ is a permutation operator interchanging the perturbations X, Y, and Z. The tensors appearing in this equation are defined in Table \ref{tab:matrix of CCQR}, with their dimensions determined by the number of excitations considered in the model (in this work CCSD, so single and double excitations relative to the reference state). 
These definitions are consistent with the ones given by ~\citet{christiansen1998response}, the main difference is that in our case these tensors require use of complex algebra whereas in non-relativistic implementations it is typically assumed that matrix representations are either real of fully imaginary. This difference is caused by the intrinsic inclusion of spin-orbit coupling effects.

\begin{table}[H]
\begin{threeparttable}

    \centering
    \setlength{\tabcolsep}{18.0mm}{
    \begin{tabular}{cc}
    \hline
     $\boldsymbol{\eta}^{Y}$      & $\bra{\Lambda}[Y,\hat{\tau}_{\mu}]\ketCC$ \\
     $\boldsymbol{\xi}^{Y}$     & $\bra{\bar{\mu}}Y\ketCC$  \\
     $\mathbf{F}$      & $\left< \Lambda \left | \left[ \left[ H_0,\hat{\tau}_{\mu} \right],\hat{\tau}_{\nu} \right] \right|\mathrm{CC}\right>$ \\
     $\mathbf{F}^{Y}$ & $\left< \Lambda \left | \left[ \left[ Y,\hat{\tau}_{\mu} \right],\hat{\tau}_{\nu} \right] \right|\mathrm{CC}\right>$ \\
     $\mathbf{G}$ & $\left< \Lambda \left | \left[ \left[ \left[ H_0,\hat{\tau}_{\mu} \right],\hat{\tau}_{\nu} \right], \hat{\tau}_{\sigma} \right] \right|\mathrm{CC}\right>$ \\
     $\mathbf{B}$      & $\left< \bar{\mu} \left | \left[ \left[ H_0,\hat{\tau}_{\nu} \right],\hat{\tau}_{\sigma} \right] \right|\mathrm{CC}\right>$ \\
     $\mathbf{A}^{Y}$      & $\left< \bar{\mu}| [Y,\hat{\tau}_{\mu}]|\mathrm{CC}\right>$ \\
    \hline
    \end{tabular}}
    \caption{Tensors required for the CC quadratic response function\tnote{a}}
    \label{tab:matrix of CCQR}
    \begin{tablenotes}
        \item[a] $\ket{CC}=e^{T_0}\ket{R}$ denote the regular CC reference wavefunction, and $\ket{R}$ is the reference state for the CC parametrization such as Hartree-Fock state. $\bra{\Lambda} = \bra{R} + \sum_{\mu}\bar{t}_{\mu}^{0}\bra{\bar{\mu}}$. $\bra{\bar{\mu}}=\bra{R}\hat{\tau}_{\mu}^{\dagger}e^{-T_0}\equiv\bra{\mu}e^{-T_0}$, where $\hat{\tau}_{\mu}^{\dagger}$ is the deexcitation operator, which is biorthogonal to excitation operator $\hat{\tau}_{\mu}$, satisfying $\left<R|\hat{\tau}_{\mu}^{\dagger}\hat{\tau}_{\nu}|R\right> = \delta_{\mu\nu}$. $\mu$ and $\nu$ indicate excited Slater determinants (comprising single and double excitations for the CCSD model).
    \end{tablenotes}
\end{threeparttable}
\end{table}

In CC theory, the similarity transformed Hamiltonian, $\bar{H} = e^{-\hat{T}}\hat{H}e^{\hat{T}}$, plays an important role in obtaining the amplitudes and their responses to external perturbations. Since $\bar{H}$ and its matrix representation $\bar{\bold{H}}$ are not Hermitian, the left response amplitudes are not just the complex conjugate of their right counterparts. According to the 2n+1 and 2n+2 rules in perturbation theory\cite{christiansen1998response}, for obtaining the quadratic response, it is necessary to solve both the left and right first-order response equations, given respectively by:
\begin{equation}
     (\bar{\bold{H}}-\omega_{X}\bold{I})\bold{t}^{X}=-\boldsymbol{\xi}^{X}
     \label{rsp-rhs}
\end{equation}
and 
\begin{equation}
     \bar{\bold{t}}^{X}(\bar{\bold{H}}+\omega_{X}\bold{I})=-\boldsymbol{\eta}^{X}-\bold{F}\bold{t}^{X}
     \label{rsp-lhs}
\end{equation}
Within the EOM-CC approximation, the quadratic response function is expressed below\cite{pawlowski_molecular_2015,coriani_molecular_2016,faber_resonant_2019}:
\begin{equation}
    \begin{split}
        ^{EOM}\langle\langle X;Y,Z\rangle\rangle_{\omega_{Y},\omega_{Z}} =& \frac{1}{2}C^{\omega}P^{X,Y,Z}\\
        &[-^{EOM}\bar{\bold{t}}^{X}(\omega_{X})\bold{t}^{Y}(\omega_{Y})\bar{\bold{t}}^{0}\boldsymbol{\xi}^{Z}\\
        &+^{EOM}\bold{\bar{t}}^{X}(\omega_{X})^{EOM}\bold{A}^{Y}\bold{t}^{Z}(\omega_{Z})\\
        &-\bar{\bold{t}}^{0}\bold{t}^{Y}(\omega_{Y})^{EOM}\bar{\bold{t}}^{Z}(\omega_{Z})\boldsymbol{\xi}^{X}]
    \end{split}
\end{equation}
where $\bar{\bold{t}}^{0}$ indicates the zeroth-order multipliers, which can be obtained by solving the ground state Lambda equations~\cite{Shee2016}, and $^{EOM}\bold{A}^{X}$ is the EOM-CC property Jacobian matrix  
\begin{equation}
    ^{EOM}\bold{A}_{\mu\nu}^{X} = \bra{\mu}\Big[\bar{X},\ket{\nu}\bra{HF}\Big]\ket{HF}
\end{equation}
\begin{equation}
    \bar{X} = e^{-\hat{T}}\hat{X}e^{\hat{T}}
\end{equation}

The EOM-CC response is known to have an identical right response equation, as indicated in equation \ref{rsp-rhs}, when compared to linear response theory. On the other hand, EOM-CC left response equation is different from equation \ref{rsp-lhs} due to an approximation of the $\bold{F}$ matrix leading to the expression:
\begin{equation}
     ^{EOM}\bar{\bold{t}}^{X}(\bar{\bold{H}}+\omega_{X}\bold{I})=-\boldsymbol{\eta}^{X}-\bar{\bold{t}}^{0}_{D}\boldsymbol{\xi}_{S}^{X}+(\bar{\bold{t}}^{0}\boldsymbol{\xi}^{X})\bar{\bold{t}}^{0}.
     \label{rsp-lhs-eom}
\end{equation}

The detailed working equations for the matrix elements of the different terms in Eqs.~\ref{rsp-rhs} and~\ref{rsp-lhs-eom} are given in our previous linear response work\cite{yuan_formulation_2023}, including those for $\boldsymbol{\sigma}$ vectors (the products $\bar{\bold{H}}\bold{t}^{X}$ and $\bar{\bold{t}}^{X}\bar{\bold{H}}$) and property gradients $\boldsymbol{\xi}^{X}$. The working equations for new terms appearing in the quadratic response functions, such as the $^{EOM}\bold{A}^{X}$ matrix are presented in the supplementary information. 

To define a two-photon absorption cross-section, we first consider the sum-over states expression for the two-photon transition matrix elements between the reference state $\ket{0}$ and the target excited state $\ket{f}$\cite{cronstrand_multi-photon_2005}:
\begin{equation}
    \begin{split}
        T_{XY}^{f0}(\omega) = \sum_{n}\left[ \frac{\bra{f}\hat{X}\ket{n}\bra{n}\hat{Y}\ket{0}}{\omega_{n}-{(\omega+i\gamma})}+\frac{\bra{f}\hat{Y}\ket{n}\bra{n}\hat{X}\ket{0}}{\omega_{n}-(\omega^\prime+i\gamma)} \right]
    \end{split}
    \label{tpa:scattering amplitude}
\end{equation}
where $\gamma$ is the damping factor representing the inverse lifetime. The frequencies $\omega$ and $\omega^\prime$ represent the two external photons, while $\omega_{f}$ corresponds to the excitation energy between reference state $\ket{0}$ and the final excited state $\ket{f}$. For TPA the relation:
\begin{equation}
    \omega+\omega'-\omega_{f} = 0
\end{equation}
should be satisfied which means that for a given final state there is only one independent variable, whether for the most commonly studied case\cite{cronstrand_multi-photon_2005} of $\omega' = \omega = \omega_{f} / 2$  or for cases in which $\omega' \neq \omega$ such as in resonant inelastic X-Ray scattering (RIXS)\cite{rehn_resonant_2017,faber_resonant_2019}. \footnote{Note that in these references the frequencies of the absorbed and emitted photon are both defined as positive, while we define the frequencies of absorbed photons as positive and also take the frequency $\omega_f$ corresponding to the excitation energy as positive. Scattering can in our implementation be studied by defining $\omega'$ as negative in the input.} Within a response formulation, the EOM-CC right and left frequency-dependent transition moments are written as~\cite{faber_resonant_2019}
\begin{equation}
\begin{split}
     \mathrm{\textbf{Right:}} \qquad ^{EOM}T_{XY}^{f0}(\omega)=&-\bold{L}_{f}[^{EOM}\bold{A}^{X}\bold{t}^{Y}(\omega+i\gamma) + ^{EOM}\bold{A}^{Y}\bold{t}^{X}(\omega'-i\gamma)\\
     &-(\bar{\bold{t}}_{0}\boldsymbol{\xi}^{X})\bold{t}^{Y}(\omega+i\gamma)-(\bar{\bold{t}}^{0}\boldsymbol{\xi}^{Y})\bold{t}^{X}(\omega'-i\gamma)\\
     &-(\bar{\bold{t}}^{0}\bold{t}^{Y}(\omega+i\gamma))\boldsymbol{\xi}^{X}-(\bar{\bold{t}}^{0}\bold{t}^{X}(\omega'-i\gamma))]
\end{split}
\end{equation}
\begin{equation}
\begin{split}
    \mathrm{\textbf{Left:}} \quad ^{EOM}T_{XY}^{0f}(\omega) = &-[^{EOM}\bar{\bold{t}}^{X}(-\omega'-i\gamma)^{EOM}\bold{A}^{Y}+^{EOM}\bar{\bold{t}}^{Y}(-\omega+i\gamma)^{EOM}\bold{A}^{X}\\
    &-(\bar{\bold{t}}^{0}\boldsymbol{\xi}^{X})^{EOM}\bar{\bold{t}}^{Y}(-\omega+i\gamma)-(\bar{\bold{t}}_{0}\boldsymbol{\xi}^{Y})^{EOM}\bar{\bold{t}}^{X}(-\omega'-i\gamma)]\bold{R}_{f} \\
    &+ (\bar{\bold{t}}_{0}\bold{R}_{f})[^{EOM}\bar{\bold{t}}^{Y}(-\omega+i\gamma)\boldsymbol{\xi}^{X}+^{EOM}\bar{\bold{t}}(-\omega'-i\gamma)\boldsymbol{\xi}^{Y}]
\end{split}
\end{equation}
where $\bold{R}_{f}$ and $\bold{L}_{f}$ are right and left target excited states, respectively, obtained by solving EOM excitation energy (EOM-EE) equations: 
\begin{equation}
    \bar{\bold{H}} \bold{R}_{f} =  E_{f} \bold{R}_{f}
\end{equation}
\begin{equation}
    \bold{L}_{f}\bar{\bold{H}} = \bold{L}_{f} E_{f} 
\end{equation}
where the operators $\hat{R}^{f}$ and $\hat{L}^{f}$ are given by in terms of the electron-creation ($a_{a}^{\dag}$ and $a_{b}^{\dag}$) and electron-annihilation operators ($a_{i}$ and $a_{j}$)
\begin{equation}
    \hat{R}^{f} = r_{0} + \sum_{ia}r_{i}^{a}a_{a}^{\dag}a_{i} + \sum_{i>j,a>b}r_{ij}^{ab}a_{a}^{\dag}a_{b}^{\dag}a_{i}a_{j}
\end{equation}
\begin{equation}
    \hat{L}^{f} = l_{0} + \sum_{ia}l_{a}^{i}a_{i}^{\dag}a_{a} + \sum_{i>j,a>b}l_{ab}^{ij}a_{i}^{\dag}a_{j}^{\dag}a_{a}a_{b}
\end{equation}

With these left and right transition moments available, the total scattering amplitudes can then be evaluated by the equation~\cite{christiansen1998response,faber_resonant_2019}:
\begin{equation}
    S_{XY,ZU}=T_{XY}^{0f}(\omega)T_{ZU}^{f0}(\omega) = \frac{1}{2} [T_{XY}^{0f}(\omega)T_{ZU}^{f0}(\omega) +  (T_{ZU}^{0f}(\omega)T_{XY}^{f0}(\omega))^{*}]
\end{equation}

Finally, the TPA cross-section, $\delta_{TPA}$, is determined by the components of scattering amplitudes matrix $\mathbf{S}$\cite{mcclain_excited_1971}:
\begin{equation}
    \delta_{TPA} = \frac{1}{15}\{F\sum_{X,Y}S_{XX,YY}+G\sum_{X,Y}S_{XY,XY}+H\sum_{X,Y}S_{XY,YX} \}
\end{equation}

The constants $F$, $G$, and $H$ depend on the polarization of the incident light. In this work, $F = G = H = 1$ is selected to represent parallel linearly polarized light. Moreover, we set up the frequency of the external field as half of the excitation energy of the target state ($\omega'=\omega=\omega_{f}/2$).

\section{Computational details}

All EOM-CC quadratic response and two-photon absorption calculations were carried out with development versions (see revision number in SI) of the DIRAC code\cite{saue2020dirac,DIRAC23}, employing the uncontracted triply-augmented valence triple zeta Dyall basis set (defined as t-aug-dyall.v3z in inputs) for heavy elements (In, I, At, Xe, Rn)~\cite{dyall2022diffuse,dyall2006relativistic}, and an equivalent triply-augmented uncontracted Dunning basis set (defined as t-aug-cc-pVTZ in inputs) for light elements (H, F, Cl, Ga, Br)~\cite{kendall1992electron,woon1993gaussian,wilson1999gaussian}. We utilized the exact two-component (X2C)\cite{iliavs2007infinite} relativistic Hamiltonian, and in some cases, to show the effect of relativity explicitly, we also provide results using the non-relativistic Hamiltonian\cite{levy1967nonrelativistic,visscher2000approximate}(as activated by the \texttt{.Levy-Leblond} keyword). To study the effect of electron correlation, we performed quadratic-response and two-photon absorption calculations based on mean-field methods such as Hartree-Fock (HF) and density-functional theory (employing the B3LYP\cite{becke1993new} density functional approximation). The relativistic and non-relativistic calculations have been carried out with the Gaussian type\cite{visscher1997dirac} and point charge nucleus model, respectively. 

In what follows, we shall use the term orbital as shorthand for both spinors and spin-orbitals, depending on the Hamiltonian used in the calculation.

In our calculations for heavy elements (HI, HAt, Xe, and Rn), we have profited from the components of an ongoing implementation in ExaCorr of the Cholesky-decomposition approach\cite{beebe1977simplifications,koch2003reduced,aquilante2007low} to reduce the memory footprint of our calculations in the step to transform two-electron integrals from AO to MO basis, with thresholds of 10$^{-9}$ (Xe and Rn), and 10$^{-4}$ (HI and HAt), the latter is looser than the one employed in our previous work; we have carried out benchmark calculations on selected systems to verify this change did not significantly alter our results.

The molecular structures employed in all calculations have been taken from the literature: from~\citet{huber1979constants} for HX (X=F, Cl, Br, I), and from \citet{pereira_gomes_influence_2004}
for HAt. The internuclear distances employed are thus H-F (0.91680 \AA), H-Cl (1.27455 \AA), H-Br (1.41443 \AA), H-I (1.60916 \AA), and H–At (1.722 \AA).

In the calculations, the size of the correlated virtual spaces in the coupled cluster is truncated by discarding orbitals with energies above 5 a.u. For the occupied orbitals,  we correlate only valence electrons.

\section{Results and discussion}

\subsection{First hyperpolarizability of HX(X=F, Cl, Br, I, At)}
To demonstrate our implementation we first apply it to calculate the parallel component of the static first hyperpolarizability ($\beta_{||}$)\cite{shelton_measurements_1994} of the hydrogen halide molecules.
\begin{equation}
    \beta_{||} = \frac{1}{5}\sum_{i=x,y,z}(\beta_{iiz}+\beta_{izi}+\beta_{zii})
\end{equation}

Each component is defined by the equation\cite{norman_principles_2018}:
\begin{equation}
    \beta_{ijk}(-\omega_{\sigma};\omega_{1},\omega_{2}) = \sum P_{-\sigma,1,2}\sum_{n,m}\frac{\bra{0}\hat{\mu}_{i}\ket{n}\bra{n}\hat{\mu}_{j}\ket{m}\bra{m}\hat{\mu}_{k}\ket{0}}{(\omega_{n0}-\omega_{\sigma})(\omega_{m0}-\omega_{2})}
    \label{fre-den-pol}
\end{equation}

\noindent where $\hat{\mu}_{i}$ are Cartesian components of the electric dipole operators, and $\sum P_{-\sigma,1,2}$ indicates the sum of six terms by permuting the pairs ($i,-\omega_{\sigma}$), ($j,\omega_{1}$), ($k,\omega_{2}$).

Before proceeding with the calculation, it is crucial to select an appropriate basis set and establish the correlation space. Our study evaluates the impact of the basis set and correlation space on the hyperpolarizability of HF molecules. The results are presented in Table \ref{tab:static polarizability of hf}  where they are compared with results from the DALTON program\cite{aidas_DALTON_2014,hattig_frequency-dependent_1997} and experimental data.

An analysis of the first three rows reveals that both diffuse functions and polarization functions significantly influence the calculation of hyperpolarizability, as is well-known in the literature\cite{hattig_frequency-dependent_1997,rizzo_coupled_2002}. For example, when utilizing the doubly-augmented d-aug-cc-pVDZ basis set, the result is only 58\% of the value obtained with the augmented s-aug-cc-pVDZ basis set. Conversely, the effect of the correlation space is relatively minor. By comparing the results of the third and fourth rows, it is evident that correlating all virtual orbitals enhances the value by merely around 1\%.

In the fourth row, we observe that our calculation, when using the non-relativistic Hamiltonian, matches the DALTON value (-7.3385 a.u.) precisely and this serves as a validation of our implementation. Furthermore, based on the DALTON results, the disparity between EOM-QRCC and QR-CC is approximately 4.5\%. This deviation stems from the absence of size extensivity in the transition moments of the EOM model. This inconsistency between EOM and CC was previously highlighted in research on linear response properties (see \citet{yuan_formulation_2023} and references therein), and we plan to delve deeper into this topic by studying a wider array of molecules for both linear and quadratic response properties in follow-up work.

To compare with experimental data we need to account for the fact that the available value (-10.88$\pm$0.95 a.u.) concerns a value measured for a frequency corresponding to 0.0656 a.u. rather than to the static limit. Taking this into account increases $ \beta_{||}$ by almost 1 a.u. to -8.79 a.u. which is still outside the experimental error bar. Beyond the limitations of the basis set, which could still be further improved, also vibrational effects will contribute to this observed discrepancy. These effects can amount to -1.24 a.u. as discussed by \citet{hansen_automated_2009} who treated these with the vibrational configuration interaction method.

Comparison between the fourth, fifth, and sixth rows of the QR-CC calculations reveals that adding more diffuse functions (from d-aug-cc-pVTZ to t-aug-cc-pVTZ) improves accuracy more significantly than incorporating additional polarization functions (from d-aug-cc-pVTZ to d-aug-cc-pVQZ). Given that the QZ calculations are notably more resource-intensive than TZ ones, we will employ the t-aug-cc-pVTZ basis set for the subsequent calculations on heavier elements.

        \begin{table}[H]
        \begin{threeparttable}        
            \center
            \caption{Benchmark of basis sets and correlation virtual orbital space for QR-CC calculations of the static $ \beta_{||}$ (a.u.) of the HF molecule}\label{tab:static polarizability of hf}

        \setlength{\tabcolsep}{7.0mm}{
            \begin{tabular}{cccccc}
            Basis &NR-EOM&  NR-EOM\tnote{a}  & NR-QR-CC\tnote{a}  & Exp\cite{dudley_measurements_1985} \\
            \hline
            s-aug-ccpVDZ\tnote{b}   &-9.4232 &  &  &\\
            d-aug-ccpVDZ\tnote{b}   &-5.5463  & &  &\\
            d-aug-ccpVTZ\tnote{b}   &-7.2677 &  &  &\\
            d-aug-ccpVTZ\tnote{c}   &-7.3385 &-7.3385 &-7.6718     &\\
            d-aug-ccpVQZ\tnote{c,d}  &     &     &-8.5816  &  \\
            t-aug-ccpVTZ\tnote{c,d}  &     &     &-8.7930  &  \\
                  &     &     &  &  -10.88$\pm$0.95\tnote{d}\\

            \hline
            \end{tabular}}
            \begin{tablenotes}
                \item[a] Calculations were performed using the DALTON program
                \item[b] Truncating the virtual orbital space at 5 a.u.
                \item[c] Correlating all virtual orbitals
                \item[d] $ \beta_{||}$ at frequency of 0.0656 a.u.
            \end{tablenotes}
        \end{threeparttable}
        \end{table}

In Table \ref{tab:static polarizability of diatomics}, the static hyperpolarizability of hydrogen halides molecules (from F to At) is displayed, in which we show the Hartree-Fock, B3LYP, and EOM-CC results for both the non-relativistic and the X2C Hamiltonian.

At Hartree-Fock level, $\beta_{zxx}$, $\beta_{zzz}$, and $\beta_{||}$ all generally exhibit an upward trend from HF to HAt in both relativistic and non-relativistic calculations. This pattern is also discernible in the correlated calculations, though the exact values vary slightly. Both CC and B3LYP results indicate that electron correlation tends to decrease the value of $\beta_{zxx}$ for all species on the series. Nevertheless, for the $\beta_{zzz}$, CC results indicate an increase in value due to electron correlation for all molecules except HF, whereas B3LYP shows the opposite pattern. This divergence between CC and B3LYP leads to discrepancies in the final $\beta_{||}$ value. For the heavier molecules, B3LYP values deviate considerably from both the Hartree-Fock and the CC ones.

Accounting for relativistic effects is, as expected, absolutely essential for systems containing heavier elements. For instance, both Hartree-Fock and CC models reveal that for HAt, the non-relativistic outcomes are approximately half of their relativistic counterparts. With the exception of HF, all three models consistently suggest that relativistic effects increase the $\beta_{||}$ values for all molecules. Furthermore, the effects of relativity on $\beta_{zxx}$ are much larger than that of $\beta_{zzz}$.

To understand the different magnitude of relativistic correction for $\beta_{zxx}$ and $\beta_{zzz}$, we can look at equation \ref{fre-den-pol}. Through spin-orbit coupling (SOC), relativity modifies the response function by shifting the location of the poles of the response functions and by introducing additional transition channels.  Without SOC, the ground state is in all cases of pure $^{1}\Sigma_{0}^{+}$ symmetry. When introducing SOC, the $^1\Sigma$ designation is no longer strictly valid, and allowed transitions are only characterized by the remaining quantum numbers, thereby yielding $0^{+}\to 0^{+}$, and $0^{+}\to 1$ transitions. The former is evidently connected to the $z$-component of the transition dipole moment, while the latter corresponds to the $x$ and $y$-components. Specifically, the $\beta_{zzz}$ component only permits contributions from transitions to $0^{+}$ states, originating from $^{3}\Pi_{0+}$, and $^{3}\Sigma_{0+}$ states. On the other hand, for the $\beta_{zxx}$ component, transitions to the $1$ state, potentially emerging from $^{3}\Pi_{1}$, $^{3}\Sigma_{1}$ and $^{1}\Pi_{1}$ states, are also allowed, with an enhanced contribution from the singlet states $^{1}\Pi$.

      \begin{table}[H]
        \begin{threeparttable}        
            \center
            \caption{Static hyperpolarizability (a.u.) of hydrogen halides HX(X=F, Cl, Br, I, At)}\label{tab:static polarizability of diatomics}

        \setlength{\tabcolsep}{5.0mm}{
            \begin{tabular}{lcccccc}
                &HF\tnote{a}  &HF\tnote{b}& B3LYP\tnote{a} & B3LYP\tnote{b}  &  CC\tnote{a} & CC\tnote{b}  \\
            \hline
            \multicolumn{7}{c}{$\beta_{zxx}$}\\
            \hline
            HF  &-0.5091   &-0.5100  &-1.4916   &-1.5565   &-1.5276  &-1.5286    \\
            HCl &2.3017    &2.3672  &-0.1127   &0.0117   &-0.0858  &-0.0157    \\
            HBr &5.2994    &5.9155  &2.8120   &3.8513   &2.6604 &3.4213    \\
            HI  &10.7728   &13.6493  &4.8719   &10.2006   &6.7316 &9.8900    \\
            HAt  &16.0854  &38.3140  &8.7719   &42.5584   &11.1347  &32.2297    \\
            \hline
            \multicolumn{7}{c}{$\beta_{zzz}$}\\
            \hline
            HF &-8.3950    &-8.4157  &-9.8253   &-9.8476   &-9.4973  &-9.4983    \\
            HCl &-11.4505    &-11.4435  &-13.1441   &-13.0321   &-10.2439  &-10.1522    \\
            HBr &-11.0481    &-11.1427  &-11.4062   &-10.8785   &-6.4784  &-5.9790    \\
            HI  &-2.9448    &-3.9634  &-6.0444   &-5.3245   &5.1093  &4.9752    \\
            HAt  &5.3664   &5.5476  &1.3711  &5.2236   &16.9624  &11.4539    \\
            \hline
            \multicolumn{7}{c}{$\beta_{||}$}\\
            \hline
            HF  &-5.6479    &-5.6614  &-7.6852   &-7.7763   &-7.5315  &-7.5333   \\
            HCl &-4.1082    &-4.0255  &-8.0218   &-7.8052   &-6.2493  &-6.1101    \\
            HBr &-0.2696    &0.4130  &-3.4693   &-1.9055   &-0.6946  &0.5181    \\
            HI  &11.1606    &14.0011  &2.2196   &9.0460  &11.1434  &14.8531  \\
            HAt  &22.5224    &49.3054  &11.3490   &54.2043   &23.5391  &45.5480  \\
            \hline
            \end{tabular}}
            \begin{tablenotes}
                \item [a] non-relativistic calculation using the Levy-Leblond Hamiltonian
                \item [b] Relativistic calculation using the X2C Hamiltonian
            \end{tablenotes}
        \end{threeparttable}
        \end{table}

\begin{figure}[H]
    \centering
    \includegraphics[width = .5\linewidth,height=6.5cm]{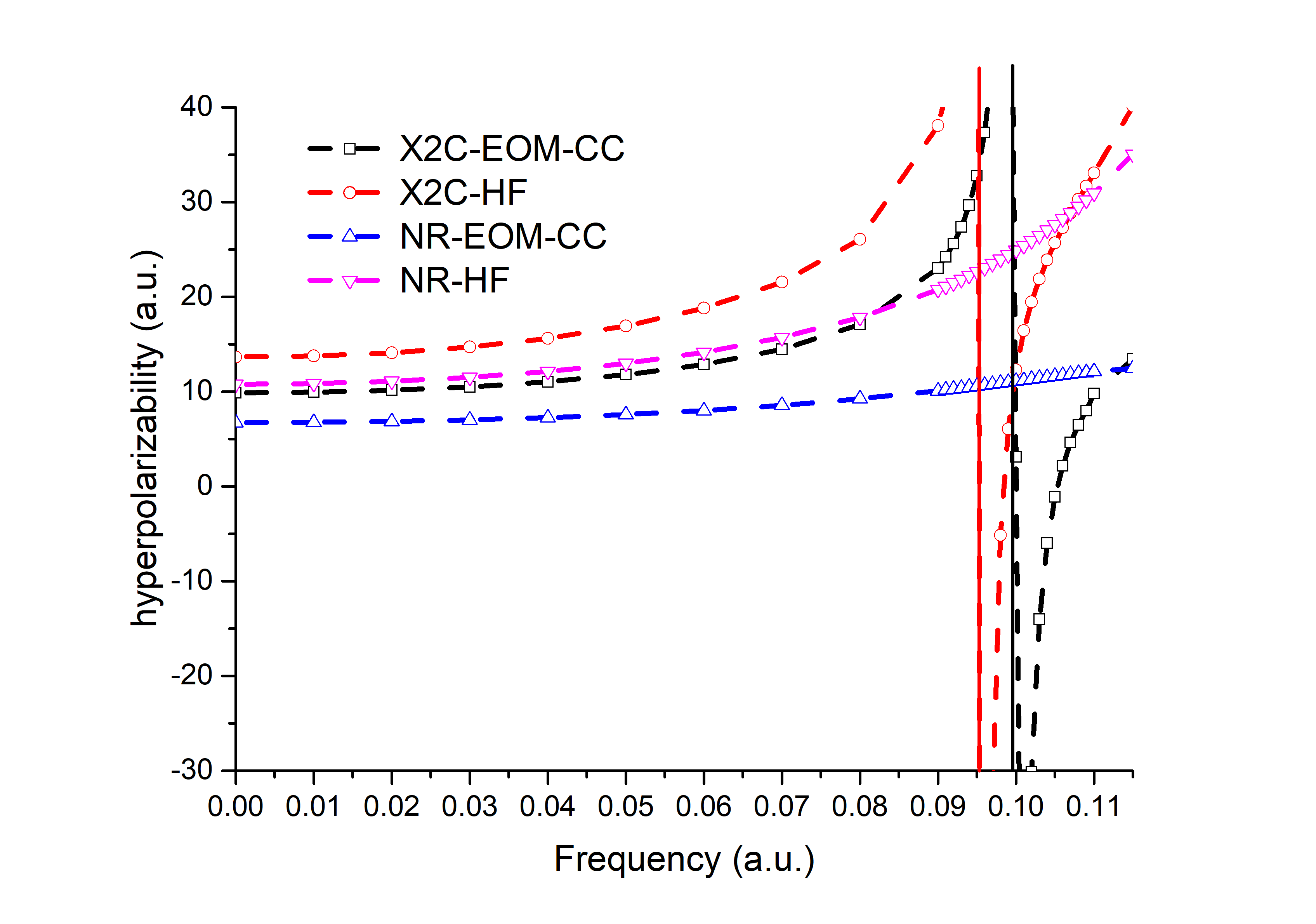}
    \includegraphics[width = .49\linewidth,height=6.5cm]{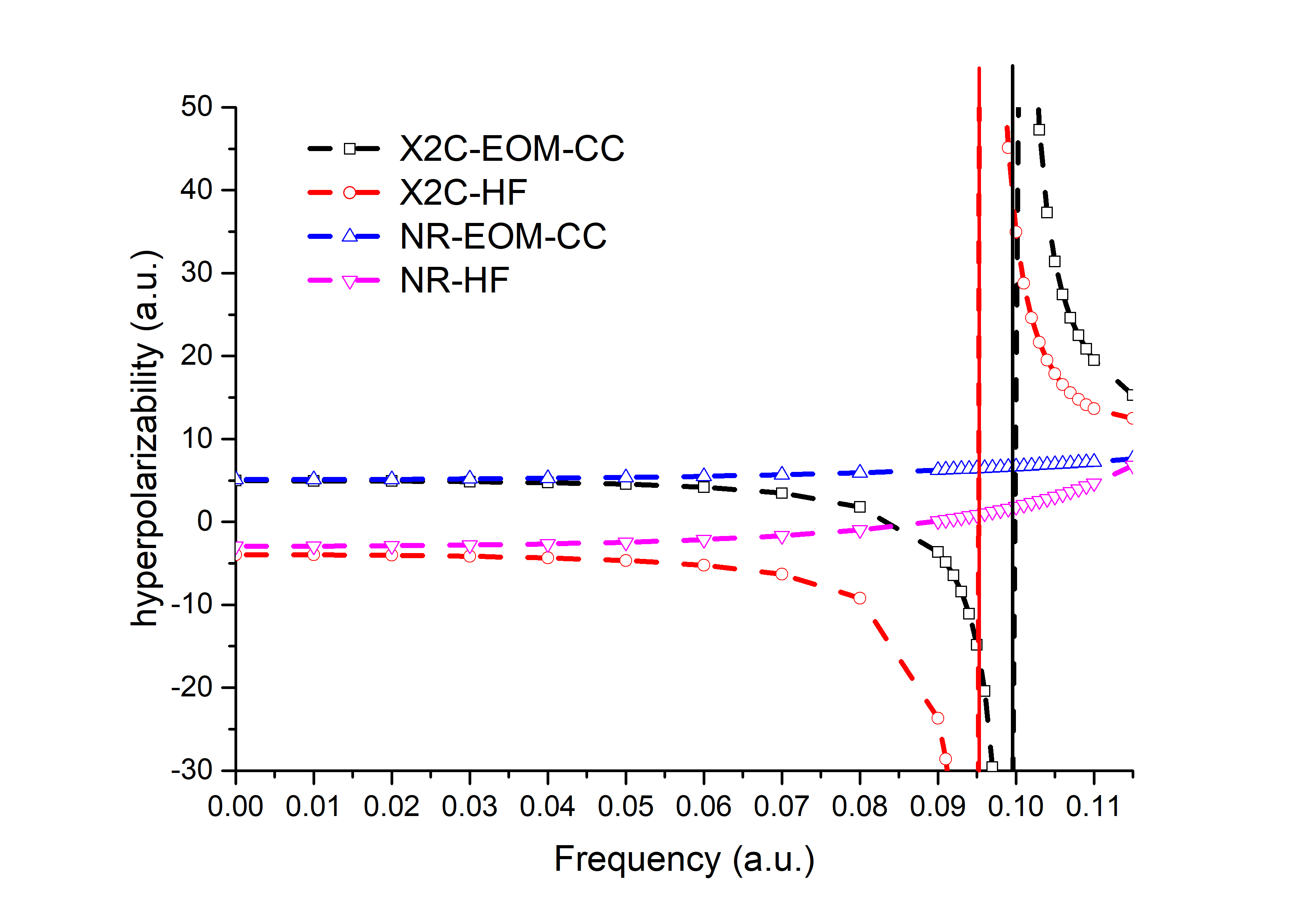}
    \caption{Frequency-dependent hyperpolarizability of HI (Left: $\beta_{zxx}$, Right: $\beta_{zzz}$). The red and black vertical lines are half of the excitation energy of $a^{3}\Pi_{0+}$ state for HF(0.0958 a.u.) and EOM-CC (0.1001 a.u.), respectively. }
    \label{fig:fre-den-hyp-hi}
\end{figure}

We now turn to the frequency-dependence of the hyperpolarizability, focusing on the impacts of relativistic and electron correlation effects. Using both CC and Hartree-Fock methods, we assess the hyperpolarizability of the single-frequency optical processes -the second harmonic generation (SHG) of HI, in which we have $\omega_{\sigma} = 2\omega_{1} = 2\omega_{2} = 2\omega$. Figure \ref{fig:fre-den-hyp-hi} displays the result of frequency ranging from 0.0 to 0.115 a.u. To further interpret these curves, we also calculate the excitation energy for the lowest five electronic states by diagonalizing $\bar{\bold{H}}$. These results are summarized in Table \ref{tab:excitation of hi}.

Overall the dispersion curves for CC and HF are qualitatively similar for both non-relativistic and relativistic calculations, though the values for CC values are larger than the HF ones, which is due to larger excitation energy of the lowest lying dipole allowed states, that can be observed in Table \ref{tab:excitation of hi}.

One can clearly find singularities in the relativistic results, located at 0.0958 a.u for HF, and located at 0.1000 a.u. for CC. According to equation \ref{fre-den-pol}, in SHG, a singularity should appear twice on the curve: the first pole corresponds to half of the excitation energy, while the second aligns with the full excitation energy. Observing this pattern, we pinpoint the singularities in our curves to the first pole associated with the $a^{3}\Pi_{0+}$ state (with excitation energies for HF and CC being 0.1916 a.u. and 0.2001 a.u., respectively).

In the absence of SOC, the transition to $a^{3}\Pi_{0+}$ is spin-forbidden. This observation aligns with the singularities appearing exclusively in the relativistic calculations. On the other hand, the pole related to the transition to $a^{3}\Pi_{1}$, is not observed, despite its permissibility with SOC. This is attributed to our focus on the $\beta_{zxx}$ and $\beta_{zzz}$ components. One can find that in equation \ref{fre-den-pol}, the transition dipole moment $\bra{0}\hat{\mu}_{z}\ket{n}$ is zero for all $\ket{n}$=$\ket{1}$ states.

        \begin{table}[H]
        \begin{threeparttable}        
            \center
            \caption{Excitation energy (a.u.) of HI for the lowest five states}\label{tab:excitation of hi}

        \setlength{\tabcolsep}{6.0mm}{
            \begin{tabular}{cccccc}
            State  &X2C-EOM-CC& X2C-HF & NR-EOM-CC  & NR-HF \\
            \hline
            $a^{3}\Pi_{2}$ &0.1773 &0.1670 &0.1894 &0.1775  \\
            $a^{3}\Pi_{1}$ &0.1831 &0.1743 &0.1894 &0.1775  \\
            $a^{3}\Pi_{0-}$ &0.1979 &0.1861 &0.1894 &0.1775  \\
            $a^{3}\Pi_{0+}$ &0.2001 &0.1916 &0.1894 &0.1775  \\
            $A^{1}\Pi_{1}$ &0.2108 &0.2120 &0.2111 &0.2189  \\

            \hline
            \end{tabular}}
        \end{threeparttable}
        \end{table}

\subsection{Verdet constant of Xe and Rn }

In the current section, we show a calculation of the Verdet constant as an illustrative example of the use of our implementation for a mixed electric-magnetic property. The Verdet constant is evaluated with the following frequency-dependent quadratic response function\cite{coriani_coupled_1997,norman_quadratic_2004}:

\begin{equation}
    V(\omega) = \omega\frac{eN\epsilon_{xyz}}{24c_{0}\epsilon_{0}m_{e}}\text{Im}\langle\langle\hat{\mu}_{x};\hat{\mu}_{y},\hat{m}_{z}\rangle\rangle_{\omega,0}
\end{equation}
with $N$ the number density of the gas, $e$ the elementary charge, $m_{e}$ the electron mass, $c_{0}$ the speed of light in vacuo, and $\hat{m}_{z}$ is magnetic dipole moment operator.

We calculate the Verdet constant at three different laser wavelengths (589 nm, 694.3 nm, and 1064 nm) for Xe and Rn and list the results in Table \ref{tab:verdet of xe}. For Xe, compared to the experiment, the HF value shows sizeable relative errors of about 10\%. The relativistic effect increases the value and reduces the error to 5\% at the HF level. It is evident that the scalar relativistic results (with the SFDC Hamiltonian) closely align with the NR-HF values, but deviate significantly from the DC and X2C results. One can note the scalar relativistic effects decrease the Verdet constant value while SOC moves the results in the opposite direction, but more strongly. This suggests that a major portion of the relativistic correction originates from the spin-orbit coupling, and considering only scalar relativistic effects may lead to an underestimation of results. Additionally, upon investigating the influence of the Gaunt interaction, we determine it to further increase the value of the Verdet constant, but much more modestly (0.5\%).

The effect of electron correlation is also to increase the Verdet constant, but the higher the degree of electron correlation recovered, the less important the increase. If we compare the DALTON NR-CCSD results obtained with a truncated correlation in space with those in which all occupied and virtual orbitals are included in the calculation, we observe a 2\% difference, with the latter calculation showing smaller values. On the other hand, comparing CCSD and CC3\cite{pecul_high-order_2006} results with the truncated correlation space, we observe a small increase in the Verdet constant for CC3. Relative to the QR-CC results, our non-relativistic  EOM calculation seems to overestimate the correlation effect by about 1\%, due to the non-extensivity issue discussed for the hyperpolarizabilities (and in ref \cite{yuan_formulation_2023}). Even though this overestimation causes the X2C-EOM value to be significantly larger than experimental results, we anticipate that, given the downward trend in the QR-CC results upon improving the correlation space discussed above (approximately -0.08 (10$^{-3}$ rad/(T m) at the 1064 nm wavelength), the X2C-EOM value with a complete orbital space are expected to come closer to the experimental values.

When we examine the Rn, we find the relativistic effect to be substantial for both HF and CC. For example, at wavelength 589 nm, the non-relativistic EOM value is 18.90 (10$^{-3}$ rad/(T m)), but the relativistic EOM value is 25.91 (10$^{-3}$ rad/(T m)). Even in the absence of experimental data for reference, such a pronounced correction underscores the importance of accounting for relativistic effects. Beyond this amplified relativistic effect, other observations for Rn align with those for Xe. This includes the dominance of the relativistic effect by spin-orbit coupling, the marginal influence of the Gaunt interaction, and a comparable magnitude of difference between EOM and QR-CC.

It is also worth noting, as reported in the supplementary material of Ref \cite{ikalainen_fully_2012}, that the performance of B3LYP is somewhat poor. It tends to overestimate the values in both non-relativistic and relativistic calculations, with errors approaching 18\% for Xe compared to the experiment.

\begin{table}[H]
\begin{threeparttable}
    \centering
    \setlength{\tabcolsep}{5.0mm}{
    \begin{tabular}{cccc}
    \hline
    Method & $\lambda$(589 nm) & $\lambda$(694.3 nm)  & $\lambda$(1064 nm)\\
    \hline
    \multicolumn{4}{c}{Xe}\\
    \hline
    NR-HF   &11.18 &7.88 &3.26 \\
    SFDC-HF\tnote{d} &11.12 &7.83 &3.24 \\
    X2C-HF  &11.61 &8.17 &3.37 \\
    DC-HF\tnote{b}   &11.61 &8.17 &3.37 \\
    DCG-HF\tnote{c}  &11.66 &8.20 &3.39 \\
    NR-B3LYP &13.62 &9.56 &3.93 \\
    X2C-B3LYP &14.60 &10.21 &4.18 \\
    NR-EOM  &12.55 &8.83 &3.65\\  
    X2C-EOM &13.10 &9.21 &3.79\\
    NR-CCSD\tnote{a} &12.39 &8.72 &3.60\\
    NR-CCSD\tnote{a *} &12.11 &8.53 &3.52\\
    NR-CC3\tnote{a} &12.46 &8.77 &3.62\\
    Exp &12.30\tnote{e} & &3.56$\pm$0.10\tnote{f} \\
    \hline
    \multicolumn{4}{c}{Rn}\\
    \hline
    NR-HF &16.80 &11.80 &4.86 \\
    SFDC-HF\tnote{d} &16.56 &11.60 &4.76 \\
    X2C-HF &23.13 &16.02 &6.48 \\
    DC-HF\tnote{b} &23.14 &16.03 &6.48 \\
    DCG-HF\tnote{c} &23.25 &16.10 &6.51 \\
    NR-B3LYP &20.08 &14.04 &5.74 \\
    X2C-B3LYP &30.38 &20.83 &8.31 \\
    NR-EOM &18.90 &13.26 &5.44\\  
    X2C-EOM &25.91 &17.92 &7.22\\
    NR-CCSD\tnote{a} &18.65 &13.08 &5.37 \\

    \hline
    \end{tabular}}
    \begin{tablenotes}
        \item[a] Calculations were performed using the DALTON program
        \item[*] Include all occupied and virtual orbitals
        \item[b] Dirac-Coulomb Hamiltonian\cite{visscher_approximate_1997}
        \item[c] Dirac-Coulomb plus Gaunt interaction Hamiltonian
        \item[d] Dirac-Coulomb without spin-orbit coupling Hamiltonian\cite{dyall_exact_1994}
        \item[e] Reference\cite{Ingersoll:56}
        \item[f] Reference\cite{cadene_circular_2015} 
    \end{tablenotes}
    \end{threeparttable}
    \caption{Verdet Constant V($\omega$) [in 10$^{-3}$ rad/(T m)] for Gaseous Xe and Rn at Different Laser Wavelengths}
    \label{tab:verdet of xe}
\end{table}

\subsection{Two-photon absorption cross-sections}

Finally, we consider the two-photon absorption cross-sections. As a showcase for our implementation, we highlight the two-photon transitions for group IIIB divalent ions, namely Ga$^{+}$ and In$^{+}$. Such systems have been discussed in the literature\cite{zuhrianda_anomalously_2012,safronova_precision_2011} for their potential use as an atomic clock. As an initial step in this exploration, we focus on spin-allowed transitions, allowing for comparative analyses with other non-relativistic programs. We intend to address spin-forbidden transitions in subsequent work.


In Table \ref{tab:tpa of iiib}, we calculate the TPA cross-sections and the corresponding excitation energy of the target states for Ga$^{+}$, and In$^{+}$ by HF and CC. Both methods indicate that relativistic effects reduce TPA cross-sections. Notably for In$^{+}$, the X2C value is roughly 60\% of its non-relativistic counterpart. When comparing HF and CC, we find that electron correlation further diminishes the TPA cross-sections. For Ga$^{+}$, the electron correlation effect is slightly larger than that of relativity, whereas, in the case of  In$^{+}$, relativistic effects show larger contributions to the final TPA cross-sections than electron correlation.

The observed trend, wherein both electron correlation and relativistic effects reduce the TPA, can be understood by examining their effect on the excitation energies. We find that both factors increase the excitation energy of the target state ($(4s5s)^{1}S_{0}$ for Ga$^{+}$, $(5s6s)^{1}S_{0}$ for In$^{+}$) and from equation \ref{tpa:scattering amplitude}, a higher excitation energy in the denominator usually results in a smaller transition amplitude. 

In non-relativistic calculations, our EOM results closely match the TPA results derived from QR-CC in DALTON, with only about a 3\% discrepancy, in line with the discrepancies observed for the hyperpolarizabilties. 

\begin{table}[H]
\begin{threeparttable}

    \centering
    \setlength{\tabcolsep}{2.5mm}{
    \begin{tabular}{cccccccc}
    \hline
    Systems(transitions)  &NR-HF &X2C-HF &DC-HF &NR-EOM\tnote{a}   & X2C-EOM\tnote{a} & NR-CC\tnote{b}\\
    \hline
    \multicolumn{7}{c}{Two-photon absorption cross-sections}\\
    \hline
    Ga$^{+}$(4s-5s) &3211.01 &2710.27 &2704.53 &2535.17  & 2128.33  &2455.97\\
    In$^{+}$(5s-6s) &7964.78 &4851.59 &4831.34 &6203.59  &3745.02  &6022.11\\
    \hline
    \multicolumn{7}{c}{Excitation energy}\\
    \hline
    Ga$^{+}$(4s-5s) &0.4382 &0.4472 &0.4473 &0.4718  & 0.4812  &0.4718\\
    In$^{+}$(5s-6s) &0.3760 &0.3983 &0.3986 &0.4042  &0.4269  &0.4042\\
    \hline
    \end{tabular}}
    \begin{tablenotes}
        \item[a] Correlate both (n-1)d and (n)s shell of total 12 electrons
        \item[b] Calculations were performed using the DALTON program
    \end{tablenotes}
    \end{threeparttable}
    \caption{Two-photon absorption cross-sections $\delta$ (a.u.) and excitation energy (a.u.) of the target states for Ga$^{+}$, and In$^{+}$}
    \label{tab:tpa of iiib}
\end{table}

 We test MP2FNO-based EOM-CC energy and TPA for Ga$^{+}$ and display the results in Table \ref{tab:mp2fno tpa}. The relativistic MP2FNOs are generated based upon our previous implementation in DIRAC\cite{yuan_assessing_2022}. From the first three columns, it becomes apparent that when using standard selection schemes, we cannot achieve reasonable TPA values in comparison to the original canonical orbital results, even with a threshold of 1.0d$^{-6}$ for the occupation number where we retrieve more than 99.99\% correlation energy for the ground state. We find the excitation energy of the TPA target state (4s5s)$^{1}S_{0}$ is markedly overestimated in FNO calculations, especially when using threshold of 1.0d$^{-4}$ and 1.0d$^{-5}$. 

At the MP2 level, the 5s orbital has a small contribution to the correlation energy in the ground state. As a result, when we obtain the MP2 density matrix and natural orbitals, the occupation numbers for natural orbitals primarily influenced by 5s orbitals are exceedingly small. These orbitals are therefore omitted by the selection scheme based purely on the threshold of occupation numbers. But while the 5s orbital is not be particularly important for the ground state, it plays a significant role in the excited state under consideration. 

Recognizing this, we've adjusted the selection scheme in the MP2FNO implementation. Besides selecting natural orbitals with occupation numbers exceeding the threshold, we also incorporate all doubly-degenerate orbitals with occupation numbers below the threshold (in atomic systems, that corresponds to s$_{1/2}$ and p$_{1/2}$ orbitals). We provide a more in-depth discussion on this point in the supplementary information, where we include the excitation energy for the eight lowest states and the energy of virtual orbitals.

With this expanded natural orbital space, we revisit the CC and TPA calculations, presenting the outcomes in the 4th to 6th columns of Table \ref{tab:mp2fno tpa}. We note a marked improvement in the excitation energy of the target state. Even at a threshold of 1.0d$^{-4}$, the discrepancy when compared to full canonical results is around 1\%. For the TPA, we also see more accurate results. For example, at a threshold of 1.0d$^{-6}$, the error drops from 57\% to 30\% by correlating only two additional doubly-degenerate orbitals. We also detect a consistent trend of approaching the canonical orbitals results when going from 1.0d$^{-4}$ to 1.0d$^{-6}$.

 There remains a 30\% discrepancy between truncated FNOs and canonical orbital results. From equation \ref{tpa:scattering amplitude}, to achieve precise scattering amplitudes, we cannot overlook the impact on the transition dipole moment, which is highly sensitive to diffuse orbitals. Conversely, these diffuse orbitals will have low occupation numbers in the MP2 calculations for the ground state and will therefore be removed from the correlating space, even with the slightly modified procedure we used.

\begin{table}[H]
\begin{threeparttable}

    \centering
    \setlength{\tabcolsep}{3.0mm}{
    \begin{tabular}{cccccccc}
    \hline
    FNO\tnote{a} &FNO\tnote{b} &FNO\tnote{c}  & FNO\tnote{d} & FNO\tnote{e}  &FNO\tnote{f}  & Canonical &Exp\tnote{g}\\
    \hline
    \multicolumn{8}{c}{Number of correlated virtual orbitals}\\
    \hline
    25 &35 &48  & 29 &38& 50& 82 &\\
    \hline
    \multicolumn{8}{c}{CCSD ground state correlation energy}\\
    \hline
    -0.2371 &-0.2396 &-0.2399  & -0.2373 &-0.2397& -0.2399& -0.2399 &\\
    \hline
    \multicolumn{8}{c}{Excitation energy of the target state (4s5s)$^{1}S$}\\
    \hline
    0.5864 &0.6414 &0.5192  & 0.4861  &0.4838& 0.4835&  0.4812 &0.4860\\
    \hline
    \multicolumn{8}{c}{Two-photon absorption cross-sections}\\
    \hline
    0.19976 &$<$1.0d$^{-15}$ &3353.67  &200.06  &1300.01& 1491.51 &2128.33 \\
    \hline
    \end{tabular}}
    \begin{tablenotes}
        \item[a] FNOs with the threshold of occupation number 1.0d$^{-4}$
        \item[b] FNOs with the threshold of occupation number 1.0d$^{-5}$
        \item[c] FNOs with the threshold of occupation number 1.0d$^{-6}$
        \item[d] FNOs with the threshold of occupation number 1.0d$^{-4}$ plus doubly-degenerate orbitals
        \item[e] FNOs with the threshold of occupation number 1.0d$^{-5}$ plus doubly-degenerate orbitals
        \item[f] FNOs with the threshold of occupation number 1.0d$^{-6}$ plus doubly-degenerate orbitals
        \item[g] Results from NIST
    \end{tablenotes}
    \end{threeparttable}
    \caption{Performance of MP2FNOs on Two-photon absorption cross-sections $\delta$ (a.u.) and excitation energy (a.u.) of the target states for Ga$^{+}$}
    \label{tab:mp2fno tpa}
\end{table}

\section{Conclusion}
In this work, we implement the relativistic Equation-of-Motion Coupled Cluster method to study the molecular quadratic response properties and two-photon absorption cross-sections. This implementation is accomplished in the GPU-accelerated coupled cluster module of DIRAC (ExaCorr), extending our previous linear response coupled cluster code \cite{yuan_formulation_2023} to solve both left and right response equations.

We have validated the implementation by assessing the purely electric properties such as static and frequency-dependent first hyperpolarizability of six hydrogen halide molecules from HF to HAt. Using a non-relativistic Hamiltonian, our code exactly reproduces the EOM-based quadratic response properties implemented in the DALTON code. Compared to Hartree-Fock and B3LYP response calculations, our relativistic EOM quadratic response calculation shows the significance of both relativistic effect and electron correlation. 

We have also investigated the Verdet constant, a mixed electric-magnetic property, for Xe and Rn with different Hamiltonian and correlation models. Both correlation and spin-orbit coupling are found to have pronounced effects. Compared to NR-QR-CC, our NR-EOM calculation overestimated the results by roughly 1\%. While the X2C-EOM calculation deviates from the experimental value more than its non-relativistic counterpart, we find this to be due to error cancellation in the treatment of electron correlation, and we estimate that using larger correlating spaces should bring our X2C results more in line with experiment. We note such calculations are currently not feasible with our single-node code due to memory limits in computational resources at our disposal.

On the other hand, as consistent with previous works, we also observe the performance of B3LYP on the Verdet constant is poor with an error of 18\% compared to the experiment for Xe. Such large deviations suggest it is important to investigate these properties with the more reliable coupled cluster method. 

At last, in our study of the two-photon absorption in Ga$^{+}$, and In$^{+}$, we find relativistic and electron correlation effects both decrease the corresponding TPA cross-sections. We analyzed the results by evaluating the excitation energy of the target state and found that both effects increase the excitation energies. 

It is worth noting that most calculations are limited in size since the quadratic response properties usually require more diffuse functions, which is challenging for the memory requirement in the current single-node implementation. There is an imperative to develop algorithms that can lower computational costs. In the current work, we utilize the MP2FNOs to reduce the virtual orbital space in TPA calculations and we find for the low-lying states, MP2FNO can effectively decrease the calculation cost while maintaining accuracy. 

For higher states such as the target state in the two-photon transition considered, the bias of MP2FNOs towards the ground state may remove diffuse orbitals which will be important for excited states. A better way to consider the influence of these diffuse orbitals is to take account of the excited state in a more sophisticated manner, such as introducing the corresponding natural transition orbitals\cite{hofener_natural_2017}. Exploring this further is among our future research objectives.  Another natural development is to extend the current code to use libraries tailored for distributed memory computing architectures, such as the ExaTENSOR library, something which we are currently pursuing.


\begin{acknowledgement}

This research used resources of the Oak Ridge Leadership Computing Facility, which is a DOE Office of Science User Facility supported under Contract DE-AC05-00OR22725 (allocations CHM160, CHM191 and CHP109). XY, LH and ASPG acknowledge funding from projects Labex CaPPA (Grant No. ANR-11-LABX-0005-01) and CompRIXS (Grant Nos. ANR-19CE29-0019 and DFG JA 2329/6-1), the I-SITE ULNE project OVERSEE and MESONM International Associated Laboratory (LAI) (Grant No. ANR-16-IDEX-0004), as well support from the French national supercomputing facilities (Grant Nos. DARI A0130801859, A0110801859, project grand challenge AdAstra GDA2210).

\end{acknowledgement}


\begin{suppinfo}

The data (input/output) corresponding to the calculations of this paper are available at the Zenodo repository under DOI: \href{http://doi.org/10.5281/zenodo.8341323}{10.5281/zenodo.8341323}.

\end{suppinfo}

\bibliography{ms}

\end{document}


\section{Working equations for CCSD EOM quadratic response}

In what follows $a,b,c,.. $ will indicate {particle lines}, $ i,j,k,... $ {hole lines}, and $ p,q,r,s,... $ general indexes~\cite{crawford2007introduction}. In all equations below we use Einstein notation. Furthermore, we define 
\begin{itemize}
    \item $P$ as a permutation operator, with : $P_{-pq} f\left(\dots pq \dots\right)= f\left(\dots pq \dots\right) - f\left(\dots qp \dots\right)$;
    \item $Y_{q}^{p} = \bra{p} Y \ket{q}$ are matrix elements of property operator $Y$.
\end{itemize}



In our implementation the property Jacobian $^{EOM}A_{\nu\mu}^{Y}$ is never formed by itself, instead we always evaluate the product of $\bar{t}^{X}$ (or $L_{f}$) and $^{EOM}A_{\nu\mu}^{Y}$. This product has the
same structure as $\eta_{\mu}^{Y}$ (Eq.~\ref{eta-y}, see definition of matrix elements in~\citet{yuan_formulation_2023} )
\begin{equation}
    (\eta_{s}^{Y})' = ^{EOM}\eta_{s}^{Y} - \bra{HF}\hat{Y}\ket{HF} \label{eta-y}
\end{equation}
\begin{equation}
    (\eta_{d}^{Y})' = \eta_{d}^{Y},
\end{equation}
with the difference of using $\bar{t}^{X}$ (or $L_{f}$) rather than $\bar{t}^{0}$ as in linear response.

Also, we reuse the $\eta_{\mu}^{Y}$ diagrams and routine by replacing the $\bar{t}^{0}$ by $\bar{t}^{X}$ to define a new intermediate $(^{XY}\eta')_{\mu}$:
%
\begin{align}
         (^{XY}\eta')_{a}^{i}=&Y_{a}^{e}(\bar{t}^{X})_{e}^{i}-Y_{m}^{i}(\bar{t}^{X})_{a}^{m}-Y_{a}^{m}(t_{m}^{e}(\bar{t}^{X})_{e}^{i})-(Y_{e}^{i}t_{m}^{e})(\bar{t}^{X})_{a}^{m}\nonumber \\
         &-1/2(t_{mn}^{fe}(\bar{t}^{X})_{fe}^{mi})Y_{a}^{n}-1/2(t_{nm}^{fe}(\bar{t}^{X})_{fa}^{nm})Y_{e}^{i}+ (\bar{t}^{X})_{ae}^{im}(\xi^Y)_{m}^{e} \\
         (^{XY}\eta')_{ab}^{ij}=&P_{-ab}P_{-ij}(\bar{t}^{X})_{a}^{i}Y_{b}^{j}-P_{-ij}(\bar{t}^{X})_{ab}^{im}Y_{m}^{j}+P_{-ab}(\bar{t}^{X})_{ae}^{ij}Y_{b}^{e}\nonumber\\
         &-P_{-ij}(t_{m}^{e}Y_{e}^{j})(\bar{t}^{X})_{ab}^{im}-P_{-ab}(t_{m}^{e}Y_{b}^{m})(\bar{t}^{X})_{ae}^{ij}
\end{align}


\section{DIRAC revision number}

\texttt{0923e70dd0}, \texttt{fe4351caf9} and \texttt{380df6b}

\section{Performance of MP2 frozen natural orbitals}

We utilize the six notations below to streamline the discussion and represent the corresponding MP2FNO selection scheme.

\begin{itemize}
    \item FNO(I): threshold of occupation number 1.0d$^{-4}$
    \item FNO(II): threshold of occupation number 1.0d$^{-5}$
    \item FNO(III): threshold of occupation number 1.0d$^{-6}$
    \item FNO(I'): threshold of occupation number 1.0d$^{-4}$ plus doubly-degenerate orbitals
    \item FNO(II'): threshold of occupation number 1.0d$^{-5}$ plus doubly-degenerate orbitals
    \item FNO(III'): threshold of occupation number 1.0d$^{-6}$ plus doubly-degenerate orbitals
\end{itemize}

From Table \ref{tab: FNO excitation} we note under the standard selection schemes: FNO(I, II, III), the excitation energy of the low-lying states: (4s4p) $^{3}P$ and (4s4p) $^{1}P$ are closed to the canonical orbitals results. On the other hand, for higher states $^{3}S$, $^{1}S$, $^{1}D$, and $^{3}D$, the excitation energy are significantly overestimated. Additionally, we note that the sequence of the excited states is incorrect for FNO(I) and FNO(II). For instance, under FNO(II), the $^{3}S$ state is positioned higher than $^{1}S$. 

This discrepancy can be traced back to the overestimation of the 5s orbital energy. As illustrated in Table \ref{tab: FNO orb ener}  when comparing the recanonicalized orbital energy of FNOs to the original 5s orbital energy (-0.0999 a.u.), the energy value of 5s orbital in FNO(I)(0.2745 a.u), FNO(II)(0.1640 a.u.), and FNO(III)(-0.0119 a.u.) are all higher. Notably, for FNO(I) and FNO(II), these are so much higher than the energies even become strongly positive.

We now shift our focus to FNO(I’), FNO(II’), and FNO(III’). We observe that the energies of the  $^{3}S$ and $^{1}S$ states are more accurate. For example, for FNO(I') the error of $^{1}S$ state is 0.0017 a.u, which is significantly less than the error observed in FNO(I) at 0.1351 a.u. This can be attributed to the fact that we achieve a notable stabilization of the 5s orbital (-0.0932 a.u.) across the FNO(I', II', III') spaces.

However, incorporating doubly-degenerate orbitals with low occupation numbers in the ground state doesn't significantly improve the $^{1}D$, and $^{3}D$ states. We also observe that the degeneracy of the components of 5p$_{3/2}$ and 4d$_{3/2}$ and 4d$_{5/2}$ orbitals in FNO(I', II', III') is sometimes broken with the $|m_j| = 1/2$ orbitals being lower than the others. This issue can likely be attributed to the scheme we employed to include the doubly-degenerate  s$_{1/2}$ and p$_{1/2}$, instead of full shells. This symmetry breaking at the orbital level is then reflected in a poorer description of the high-lying states involving those orbitals. 

We note no such thing takes place for the 4p orbitals since these have large enough occupation numbers at the ground state to always be included in the correlation treatment.

\begin{table}[H]
\begin{threeparttable}

    \centering
    \setlength{\tabcolsep}{1.0mm}{
    \begin{tabular}{ccccccccc}
    \hline
    Excited state&FNO(I) &FNO(II)  &FNO(III)  & FNO(I') & FNO(II')  &FNO(III')  & Canonical &Exp\tnote{a}\\
    \hline
    (4s4p)$^{3}P_{0}^{u}$&0.2310 &0.2126 &0.2120  &0.2117  &0.2119& 0.2113 &0.2112 &0.2158\\ 
    (4s4p)$^{3}P_{1}^{u}$&0.2333 &0.2146 &0.2140  &0.2189  &0.2141& 0.2135 &0.2132 &0.2179\\ 
    (4s4p)$^{3}P_{2}^{u}$&0.2384 &0.2188 &0.2181  &0.2383  &0.2187& 0.2181 &0.2173 &0.2221\\ 
    (4s4p)$^{1}P_{1}^{u}$&0.2657 &0.3280 &0.3249  &0.3543  &0.3271& 0.3241 &0.3221 &0.3221\\ \hline
    (4s5s)$^{3}S_{1}^{g}$&0.5981 &0.6425 &0.4915  &0.4666  &0.4647& 0.4645 &0.4630 &0.4691\\ 
    (4s5s)$^{1}S_{1}^{g}$&0.5864 &0.6414 &0.5192  &0.4861  &0.4838& 0.4835 &0.4812 &0.4860\\ \hline
    (4p$^{2}$)$^{1}D_{2}^{g}$\tnote{*}&0.5918 &0.6453 &0.5183  &0.5469  &0.5466& 0.5454 &0.4914 &0.4908\\ 
    (4s4d)$^{3}D_{1}^{g}$&0.5987 &0.6632 &0.5439  &0.5678  &0.5467& 0.5460 &0.5118 &0.5186\\
    \hline
    \label{tab: FNO excitation}
    \end{tabular}}
    
    \begin{tablenotes}
        \item[a] Results from NIST
        \item[*] 44\% from the configuration 4s4d
    \end{tablenotes}
    \end{threeparttable}
    \caption{Performance of MP2FNOs on the excitation energy (a.u.) of eight excited states for Ga$^{+}$}
    \label{tab: FNO excitation}
\end{table}

\begin{table}[H]
\begin{threeparttable}

    \centering
    \setlength{\tabcolsep}{1.5mm}{
    \begin{tabular}{cccccccc}
    \hline
        FNO(I') &FNO(I')  &FNO(II)  & FNO(II') & FNO(III)  &FNO(III')  & Canonical & Virtual orbitals  \\ \hline
        -0.1317 & -0.1955 & -0.1892 & -0.1955 & -0.1892 & -0.1955 & -0.1957 & 4p$_{1/2}$\\ 
        -0.1253 & -0.1253 & -0.1854 & -0.1854 & -0.1854 & -0.1854 & -0.1924 & 4p$_{3/2}$\\ 
        -0.1253 & -0.1253 & -0.1854 & -0.1854 & -0.1854 & -0.1854 & -0.1924 & 4p$_{3/2}$\\ \hline
        0.2745 & -0.0932 & 0.1640 & -0.0932 & -0.0119 & -0.0932 & -0.0999 & 5s$_{1/2}$\\ \hline
        0.4708 & -0.0384 & 0.1693 & -0.0384 & 0.1270 & -0.0384 & -0.0659 &5p$_{1/2}$\\ \hline
        0.4708 & 0.0579 & 0.1693 & 0.0579 & 0.1270 & 0.0579 & -0.0654 &5p$_{3/2}$\\ 
        0.4734 & 0.1870 & 0.2745 & 0.1693 & 0.1276 & 0.1270 & -0.0654 &5p$_{3/2}$\\ \hline
        0.4734 & 0.4708 & 0.4708 & 0.1693 & 0.1276 & 0.1270 & -0.0566 &4d$_{3/2}$\\ 
        0.4734 & 0.4708 & 0.4708 & 0.1870 & 0.1276 & 0.1276 & -0.0566 &4d$_{3/2}$\\ \hline
        1.6174 & 0.4734 & 0.4734 & 0.4708 & 0.1640 & 0.1276 & -0.0565 &4d$_{5/2}$\\ 
        1.6509 & 0.4734 & 0.4734 & 0.4708 & 0.1693 & 0.1276 & -0.0565 &4d$_{5/2}$\\ 
        1.6509 & 0.4734 & 0.4734 & 0.4734 & 0.1693 & 0.1693 & -0.0565 &4d$_{5/2}$\\ \hline
    \end{tabular}}
    
    \end{threeparttable}
    \caption{Orbital energy (a.u.) of lowest 12 virtual orbitals for Ga$^{+}$}
    \label{tab: FNO orb ener}
\end{table}

\clearpage

\bibliography{ms}